\documentclass[useAMS,usenatbib]{mn2e}
\usepackage{graphicx}
\usepackage[update,prepend]{epstopdf}

\title[Circular Polarization]{Source of Circular Polarization in Radio Pulsars}
\author[P. B. Jones]{P. B Jones\thanks{E-mail:
peter.jones@physics.ox.ac.uk}  \\
University of Oxford, Department of Physics, Denys Wilkinson Building,\\
Keble Road, Oxford OX1 3RH, U.K.}

\begin{document}

\date{}

\pagerange{\pageref{firstpage}--\pageref{lastpage}} \pubyear{}

\maketitle

\label{firstpage}

\begin{abstract}

It is known that the concept of limiting polarization introduced seventy years ago by K G Budden has the capacity to explain the magnitude of circular polarization seen in normal pulsars with light-cylinder radii of the order of $10^{9-10}$ cm under the assumption of a high-multiplicity electron-positron plasma.  However a review of limiting polarization under the same assumption in millisecond pulsars indicates that it is inapplicable there because the region of limiting polarization lies far outside the light cylinder.  The present paper, using the ion-proton model, evaluates circular polarization both generally, and specifically for J2144-3933, and gives a fairly detailed understanding of the observations in normal pulsars including the change of sign as a function of frequency seen in J0908-4913.  But it also fails to explain circular polarization in millisecond pulsars owing to the smaller particle number densities and birefringence of the magnetosphere in these objects.  However, the review of limiting polarization finds that, within the ion-proton model, this distinct process can describe their circular polarization. It is argued that certain features of millisecond pulsar Stokes profiles are clearly consistent with limiting polarization.
\end{abstract}


\section{Introduction}

The observation of radio pulsar polarization and the longitude variation
of its position angle more than half a century ago gave immediate support to the polar-cap model of emission (Radhakrishnan \& Cooke 1969).  However,the origin of the circular component of polarization remains enigmatic.  Whilst single-pulse studies of bright pulsars (Karastergiou et al 2003) showed that the Stokes parameter $V$ could change sign from pulse-to-pulse, the integrated values of $V$ were finite, though smaller than the linear polarization Stokes $L=(Q^{2} + U^{2})^{1/2}$, and of constant sign. Generally, circular polarization is assumed to be formed between the source, loosely defined as the surface of last absorption, and the observer, but within the light-cylinder radius $R_{LC}$.

Many authors have re-introduced the concept of limiting polarization (Budden 1952) as the source of circular polarization.  We refer to, among others, Melrose \& Stoneham (1977), Cheng \& Ruderman (1979), Barnard (1986), Lyubarskii \& Petrova (1999) and Beskin \& Philippov   (2012). Without exception these authors assume the radiation propagates through a magnetized relativistic electron-positron plasma with Lorentz factors $\gamma_{e} < 10^{2}$ and multiplicities $n_{ep} \sim 10^{1-4}$ per primary electron.  Its birefringence, here the difference $\delta n_{O} - \delta n_{E}$ between the ordinary and extraordinary mode refractive indices of the plasma, is large enough for the coherence of the O- and E-modes existing at source to be lost as the modes propagate on separate paths through the inhomogeneous plasma within the light cylinder.  The radius of limiting polarization $r_{p}$ is defined as that at which the O- and E-mode beat-frequency wavelength equals the scale length in which neutron-star rotation causes significant rotation of the ${\bf k}\times{\bf B}$ plane defining the O- and E-modes,
where ${\bf B}$ is the local magnetic flux density (see, for example, Lyubarskii \& Petrova 1999).
With the exception of Beskin \& Philippov all authors introduce $r_{p}$ as a parameter in their circular polarization calculations.  In this region there can be sufficient mixing of O- and E-modes, principally through variation of the ${\bf k} \times {\bf B}$ plane, to generate from an initially plane-polarized wave the observed orders of magnitude of $V$. Beskin \& Philippov, in a more complete development avoiding the use of $r_{p}$, find that values $V \approx 0.2$ are possible for values $n_{ep} \approx 10^{3}$  in normal pulsars, but consider millisecond pulsars (MSP) only briefly and predict much smaller circular polarization of the order of $V \approx 0.01$.

However, recent observations by Dai et al (2015) produced a very complete set of $V$ and $|V|$ measurements for 24 MSP at three frequencies. In many cases $|V| > 0.1$ and the whole set appears broadly not different from normal pulsar values, indicating some discordance with the limiting polarization predictions.

The significance of birefringence in a magnetized relativistic plasma is as a diagnostic of particle mass and number density.  Arguments have been published previously (see Jones 2020; also references cited therein) that an ion-proton magnetosphere is present in pulsars with positive polar-cap corotational charge density. In a given pulsar, the magnetospheric birefringence would be five orders of magnitude smaller than in the canonical electron-positron case.  It would be remarkable if this difference did not have some impact on observable quantities, in particular circular polarization.  Section 2 of the present paper compares limiting polarization in these two different magnetospheres with particular consideration of the MSP. It also serves to place the work described in Section 3 in context.

The ion-proton plasma has small birefringence and in consequence, coherence of the O- and E-modes at the source is maintained until $\delta n_{O} - \delta n_{E}$ is so small that the polarization becomes fixed at the observed values.  In normal pulsars, this occurs at quite small radii, $\eta \sim 10$, where $\eta$ is the polar-coordinate
radius in units of the neutron-star radius $R$.  There is negligible aberration in normal pulsars at this radius. Thus the magnetosphere behaves essentially in the same way as the quarter-wave plate of classical optics, the phase lag of the O-mode relative to the E-mode being of the order of $\pi$ or less.  We have noted previously (Jones 2016) that this is consistent with the high-resolution $V$-profiles observed by Karastergiou \& Johnston (2006).  Replacing the ion-proton system by an electron-positron plasma would result in a phase lag so large in the region of O- and E-mode separation by refraction that the orientation of the polarization ellipse would be a rapidly varying function of observed longitude owing to the inevitable longitude variation of particle Lorentz factors and hence of $\delta n_{O}$.

The paper cited previously (Jones 2016) was deficient in not adequately citing and discussing the work of the limiting polarization authors.  It also failed to develop the ion-proton circular polarization to the extent of comparison with observational characteristics.  Furthermore, it referred, incorrectly, to circular polarization in the MSP as requiring no special discussion.  For these reasons, Section 3 describes how the ion-proton system produces circular polarization but with the assumption that at emission, there is a distribution of the angle $\chi$  subtended by the radiation and the local magnetic field in the frame of the rotating neutron star. This is regarded as essential because in almost any assumption about the nature of the magnetosphere, the Lorentz factors of particles are no more than moderate,  $\gamma_{e}$ or $\gamma_{ion,p} < 10^{2}$, and $\chi$ is of the same order as the angular width of the open magnetosphere. Section 4 gives our conclusions.

\section{Limiting polarization}

The definition of $r_{p}$ is the more convenient, though not the optimum, way of comparing limiting polarization in ion-proton and electron-positron magnetospheres for both normal pulsars and MSP and here follows the work of Lyubarskii \& Petrova (1999).
The refractive index for the O-mode of a relativistic plasma in the high-field limit is $1 + \delta n_{O}$ where,
\begin{eqnarray}
\delta n_{O} = - \frac{\omega_{p}^{2}\sin^{2}\theta_{k}}
{2\gamma^{3}\omega^{2}(1 - \cos\theta_{k} +1/2\gamma^{2})^{2}}
\end{eqnarray}
for angular frequency $\omega$.  The angle subtended by ${\bf k}$ and ${\bf B}$ is $\theta_{k}$, the plasma frequency is $\omega_{p}$ and $\gamma$ the particle Lorentz factor. The E-mode increment $\delta n_{E}$ is negligible in the high-field limit.  We adopt a scale length for the variation of the magnetospheric properties $\theta_{k}$ and the spatial orientation of ${\bf k}\times{\bf B}$, which is linear in the rotation period $P$ and equal to $\epsilon R_{LC}$, where $R_{LC}$ is the light-cylinder radius and $\epsilon$ is here simply a numerical factor, $0 < \epsilon < 1$.  The value of $r_{p}$ is given by the condition $\omega \delta n_{O}\epsilon R_{LC} = c$.  Its ratio to the scale length is
\begin{eqnarray}
\frac{r_{p}}{\epsilon R_{LC}} = \frac{2\pi R}{c}\left(\frac{{\rm e}}
{c\omega}\right)^{1/3}\left(\frac{NB}{mP^{3}}\right)^{1/3}
\\   \nonumber
\left(\frac{1}{\epsilon^{2}\gamma^{3}}
\frac{\sin^{2}\theta_{k}}{(1 - \cos\theta_{k} + 1/2\gamma^{2})^{2}}\right)^{1/3}
\end{eqnarray}
in which $m$ is the plasma particle mass and $B$ is here the polar-cap value.  The number $N =1$ for an ion-proton plasma or $2n_{ep}$ for a plasma with pair multiplicity $n_{ep}$.
The central of the three brackets in equation (2) contains the quantities that change in the comparison made here: the remaining terms are regarded as common factors.  A normal pulsar is assumed to have
$B = 10^{12}$ G, $P = 1$ s; and an MSP $B = 2\times 10^{8}$ G, $P = 4$ ms.  The common Lorentz factor is $\gamma = 20$, and $\epsilon = 0.2$.  The pair multiplicity is $n_{ep} = 10^{2}$. The 
evaluations of equation (2) given in Table 1 are not rapidly varying functions of these common parameters. They are also slowly varying functions of $\theta_{k}$ for which the values assumed are multiples of $\gamma^{-1}$. Normal pulsar values assumed in Table 1 for an electron-positron plasma are, of course, consistent with a region of limiting polarization inside the light cylinder but not at radii close to the emission region (see Lyubarskii \& Petrova 1999, Beskin \& Philippov 2012).  However, the MSP values indicate that limiting polarization occurs well outside the light cylinder so that the O-mode would be expected to accommodate adiabatically to the local orientation of ${\bf B}$ within the light cylinder. Substitution into the approximate expression for Stokes $V$ found by Lyubarskii \& Petrova in this limit, gives $V \approx 0.2r_{p}/\epsilon R_{LC} \approx 0.01$, consistent with the example calculated by Beskin \& Philippov, but not with the set of experimental values of Dai et al (2015) which broadly differ little from those for normal pulsars. For a comparison, we refer to the group of 17 normal pulsars observed with high resolution and good signal-to-noise ratio by Karastergiou \& Johnston (2006).

\begin{table}
\caption{Evaluations of the limiting polarization radius $r_{p}/\epsilon R_{LC}$ given by equation (2) at $1.4$ GHz for the parameters listed in the text following that equation are shown as functions of $\theta_{k}$. They are for both normal pulsars and MSP and for each of the two magnetospheres.}

\begin{tabular}{@{}llcccc@{}}
\hline  
pulsar  & magnetosphere &  $\sin \theta_{k}$       &        &        &\\
\hline
		&  				&  0.05 &  0.1   &    0.2     & 0.4	\\  
\hline
normal   &   i-p     &	0.027   &  0.024    &  0.017  &  0.010  \\
normal   &  $e\pm$	&	1.96    &  1.70   &  1.20      &  0.75  \\
MSP      &   i-p    &	0.39    &  0.34     &  0.24    &  0.15  \\
MSP      &   $e\pm$  &  29   &  25  &  17.5   &  11   \\
\hline
\end{tabular}
\end{table}

Values of $r_{p}/\epsilon R_{LC} \ll 1$ for normal pulsars with an ion-proton plasma are so small that the circular polarization arising from field re-orientation is considerably less significant than that caused by the O-mode phase lag (Jones 2016) which is evaluated in Section 3.
The phase lag is a linear function of $BP^{-1}$ and, because for the typical MSP observed by Dai et al, the value of this quantity is smaller by a factor of $\approx 0.05$ than for normal pulsars, the phase lag derived from it is too small in MSP to be significant. 

MSP values in the ion-proton case are $r_{p}/\epsilon R_{LC} < 1$ and according to the Lyubarskii \& Petrova estimate $V \approx 4r_{p}/\epsilon R_{LC}$ appears broadly consistent with the values of $V$ found by Dai et al.  The obvious question is whether there are features of MSP Stokes $V$-profiles that are consistent with the predictions of limiting polarization.  This is the case and is addressed in Section 4.

\section{Normal pulsar circular polarization profiles}

Normal pulsars have a wide range of values of $V$ and of profile forms.
In this section, we attempt to show how this happens by calculating profiles of $V$ for a number of model polar caps and for a specific pulsar; J2144-3933.

We make the following assumptions.

(i)  The source is at a unique altitude of radius $\eta_{e}$ above the polar cap. This is also assumed to be the surface of last absorption. It has linear polarization position angle locally parallel with the projection of ${\bf B}$, a dipole field, on to a plane perpendicular to the magnetic axis as in the rotating vector model (RVM). Radiation passes through a magnetized relativistic plasma consisting of outward moving particles in the open magnetosphere at $\eta > \eta_{e}$ and thence to the observer.

(ii)  Particle motion is parallel with the local ${\bf B}$ but radiation is emitted, in the rest frame of the rotating neutron star, on cones of semi-angle $\chi$ with an intensity distribution,
\begin{eqnarray}
f(\chi) = \left(1 + \gamma^{2}\chi^{2}\right)^{-3-\alpha}
\end{eqnarray}
consistent with a Lorentz transformation from an outward-moving isotropic source plasma having a rest-frame spectral index $\alpha$.

(iii) Birefringence of the plasma at $\eta > \eta_{e}$ is the origin of the circular polarization, but $\delta n_{O}$ is so small that there is negligible differential refraction and the O- and E-modes remain coherent until birefringence is negligible.

Assumptions (ii) and (iii) are not part of the RVM.  Aberration is negligible for normal pulsars at $\eta_{e}$ and the effects of rotation appearing in the concept of limiting polarization are also negligible.
It is then possible to assume that $\delta n_{O}$ is a function only of $\eta$, with angle $\theta_{k} = \gamma^{-1}$.  Then equation (1) is replaced by,
\begin{eqnarray}
\delta n_{O} = - \frac{\omega_{p}^{2}(\eta)}{2\gamma\omega^{2}},
\end{eqnarray}
and the O-mode phase lag in the limit $\eta \rightarrow \infty$ is,
\begin{eqnarray}
\phi = \frac{R\omega{_p}^{2}(1)}{4c\gamma\omega\eta_{e}^{2}} =  
136\frac{B_{12}}{\gamma P \eta^{2}_{e}} \hspace{3mm}
{\rm rad},
\end{eqnarray}
evaluated for protons and for a frequency $1.4$ GHz.

\begin{figure}
\includegraphics[trim=30mm 40mm 40mm 60mm, clip, width=84mm]
{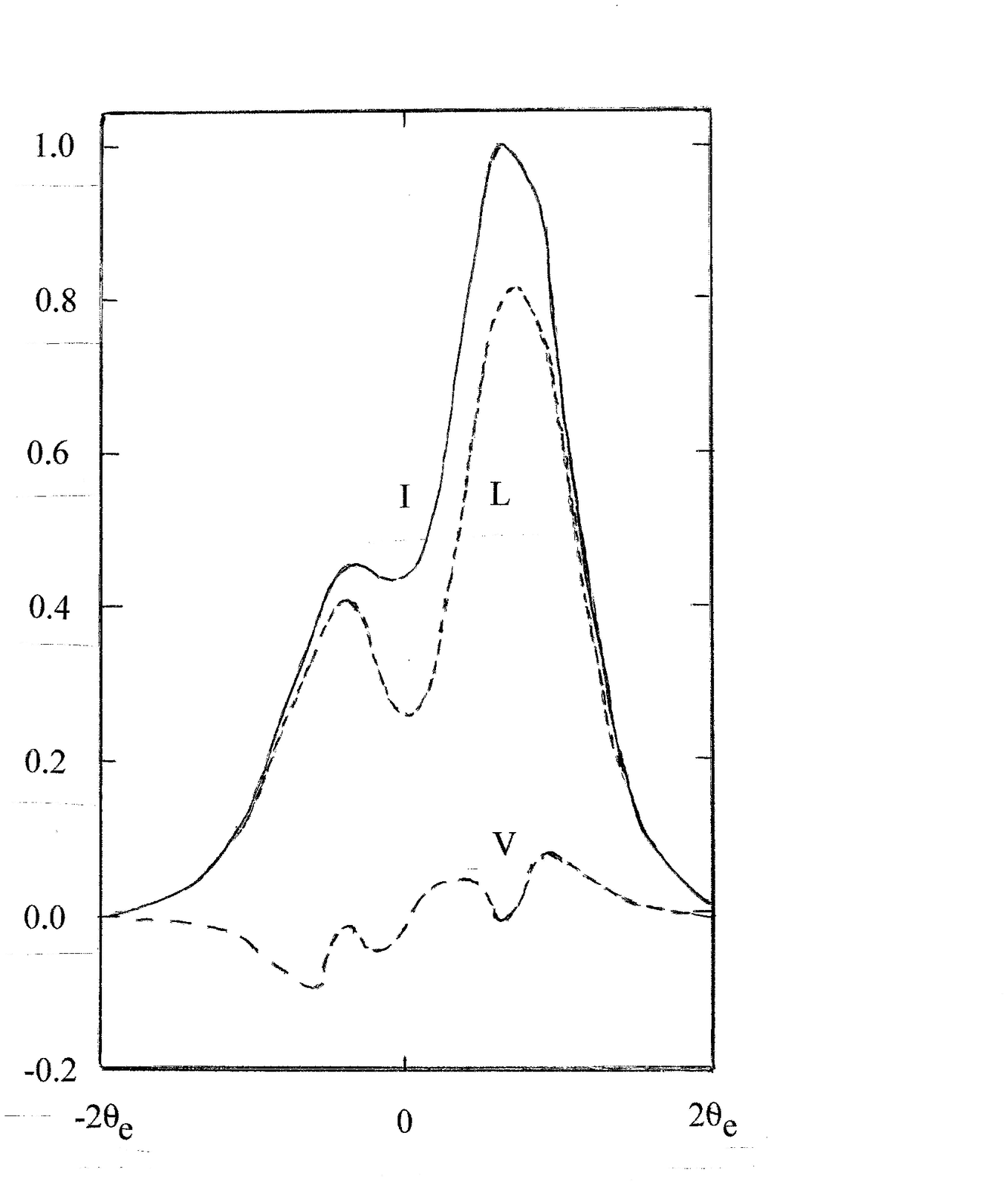}

\caption{Stokes parameters I, $L=(Q^{2} + U^{2})^{1/2}$, $V$ are shown for a $4$ $\delta$-function polar-cap intensity distribution in a general case with $P = 1$ s, $B_{12} = 1$.  They are functions of longitude $k_{\perp}\cos \zeta_{k}$ in the interval $(-2\theta_{e}, 2\theta_{e})$, for $\gamma = 20$ and $\eta_{e} = 5$, and are typical what is observed in a single pulse. The positions of the $\delta$-functions are chosen to represent a distribution consistent with conal emission.}
\end{figure}

Fig.A1 outlines the geometrical arrangement fixed in the rest frame of the rotating neutron star.  Assumption (ii) means that the unit wave-vector ${\bf k}$ which denotes the line of sight and passes from right to left on the arc of traverse, represents at any instant radiation from a finite section of the emission surface at $\eta_{e}$.  We assume that radiation from any element of the emission surface is incoherent with respect to all others.  Consequently for a given position of ${\bf k}$, we calculate and then sum the Stokes parameters for each element of area.  Fig.A1 represents in a plane perpendicular to the magnetic axis the angular deviation of flux lines from the magnetic axis at radius $\eta_{e}$.  Thus ${\bf k} = {\bf k_{\perp}} + {\bf k}_{z}$ and the local ${\bf B} = {\bf B_{\perp}} + {\bf B}_{z}$.  We make a small angle approximation so that $k_{z} =1$ and, consistently with positive corotational charge density, $B_{z} = -B$, and $B_{\perp} = -B\theta$, where $\theta$ is the flux-line deviation from the magnetic axis. Hence the vectors shown on the diagram are ${\bf k_{\perp}}$ and ${\bf B_{\perp}}$.  Following the RVM, the radiation from each element of the emission surface is assumed to be linearly polarized with position angle given by ${\bf B_{\perp}}$.

\begin{figure}
\includegraphics[trim=30mm 45mm 40mm 60mm, clip, width=84mm]{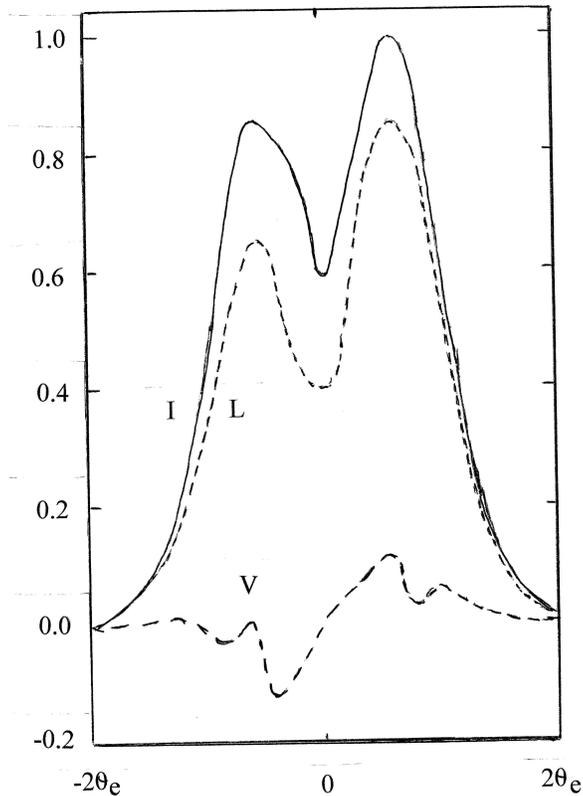}

\caption{Stokes I, $L$, $V$ are shown as in Fig.1 except that the number of $\delta$-functions has been increased to $12$ which more nearly represents a conal polar cap with most of its area active except near the magnetic axis.}
\end{figure}

The special case of a circular polar cap and dipole field is assumed only as a framework for the calculation and we can see (from equation A12 in the Appendix) that this results in zero circular polarization if the distribution of intensity on the emission surface at  $\eta_{e}$ has axial symmetry.  But the distribution of intensity over the surface at $\eta_{e}$ can be varied as appropriate for both single-pulse and integrated profiles.

It is not possible to proceed further without consideration of the form of the polar cap.  Modelling made  within the basic axial symmetry assumption (Jones 2020) shows that the conditions necessary for growth of a Langmuir mode and the development of non-linearity and turbulence do not necessarily exist over the whole polar cap at any instant of time.  These are that plasma accelerated from an element of surface area must contain both ions and protons.  The mode growth exponent is $\propto \omega^{1/2}_{p}(\eta)\gamma^{-3/2}$ so that development of turbulence is possible only for limited values of $\eta$ and, more significantly, of $\gamma$.  The critical value of $\gamma$ depends on $\omega_{p}$ but is certainly $\gamma_{c} < 10^{2}$.  The time interval within which a turbulent state can be maintained is determined by $\tau_{p}$, of the order of $1$ s, the characteristic time for a proton to diffuse from formation to the top of the neutron-star LTE atmosphere.

The geometrical shape of the polar cap in normal pulsars is unknown.
But it is known that integrated Stokes I profiles are broadly long-term stable at a given frequency although their forms are often distinctly odd, and we are driven to the conclusion that polar caps, and specifically the areas satisfying ion-proton model conditions, are themselves {\it sui generis}.  There are plausible reasons for this.
Firstly, even if circular symmetry existed at the neutron-star surface it does not do so in the vicinity of the light cylinder. This was pointed out many years ago by Arons \& Scharlemann (1979) and is consistent with force-free studies (Bai \& Spitkovsky 2010).  Recent work on J1906+0746 (Desvignes et al 2019) has been used to question this idea but must be subject to some reservations.  The magnetic axis of this pulsar is described as 9.4 degrees away from orthogonality with the rotation axis, and although emission has been seen from both sides of the pole, on one side it is weak. Secondly, much work on the evolution of the internal magnetic field involving Hall drift has indicated that complex and small-scale field components can form in the immediate vicinity of the poles. (Geppert, Gil \& Melikidze 2013; Geppert \& Vigano 2014; Wood \& Hollerbach 2015; and references cited therein.)

Guided by Karastergiou et al (2003) we consider single-pulse profiles first.  The intensity distribution within a circle of radius $\theta_{e}$ at $\eta_{e}$ can be represented by a small number of $\delta$-functions each representing the flux of particle in a turbulent state on a specific bundle of flux lines. The system the $\delta$-function represents is one of Langmuir mode growth inside an area of polar cap large enough to sustain it.  There appears to be no existing work on defining this minimum area for growth of what is usually treated as a one-dimensional mode.  Therefore, we assume that it must be at least a small multiple of $\lambda^{2}_{M}$, where $\lambda_{M} \approx 5\times  10^{2}(P/B_{12})^{1/2}\eta^{3/2}$ cm is the mode wavelength in the observer frame.  Thus it is possible that a polar cap with radius given by equation (A1) could accommodate several such systems of mode growth at any instant.  For this  reason, we believe that the $\delta$-function representation is valid and realistic for single pulses.
The polarization position angle is in the plane of the flux line at which the $\delta$-function is positioned. Results of calculations of Stokes parameters can be summarized succinctly: by choosing positions and intensities for a small number of $\delta$-functions almost any single-pulse profile can be produced including profiles of Stokes $V$ with zeros or without zeros and of either sign.  An example is shown in Fig. 1. For general cases, we adopt a spectral index $\alpha = 1.8$.

Increasing the number of $\delta$-functions produces an intensity distribution approaching that of a continuum.  Fig. 2 shows an example of this.  Karastergiou et al observed considerable variation in single-pulse profiles of a given pulsar, including changes of sign in $V$.  Integrated profiles had stable though smaller values of $\langle V\rangle$ but smaller than $\langle|V
|\rangle$  owing to the sign changes referred to.  Distinct differences between $\langle V\rangle$ and $\langle|V|\rangle$ are also present in the MSP Stokes $ V$ measurements of Dai et al (2015).  The model polar caps of the present paper are consistent with this behaviour.

Apart from the general cases, a small number of specific pulsars are worthy of consideration.  Firstly, J2144-3933 (Young, Manchester \& Johnston 1999) has $P =8.5$ s and the lowest value of $BP^{-2}$, which determines acceleration potential, of any ATNF listed pulsar (Manchester et al 2005).  Stokes parameter profiles were observed by Manchester, Han \& Qiao (1998) at $660$ MHz and are approximately: I, full width at half maximum FWHM $= 2.5$ degrees; $L/I$ double-peaked, max. $0.2$, min. $0.1$, max. $0.3$; $V/I$, min. $-0.15$, $0.00$, max. $0.2$. But their FWHM has to be reduced by a factor of $3$ owing to the re-evaluation of the period by Young et al.  The values of $u_{0}(1)$ and $\theta_{e}$ obtained from equations (A1) and (A2) are so small that a $\delta$-function representation would not be appropriate. Firstly, the ratio $\lambda_{M}/u_{0}(1)$ is large enough to indicate that no more than a single $\delta$-function can be accommodated.  Furthermore, a single $\delta$-function at the origin cannot represent the polarization of the source at $\eta_{e}$. A uniform continuum  within the circle of radius $\theta_{e}$ has been adopted but it necessarily has no circular symmetry and is shaped to be consistent with Arons \& Scharlemann (1979).
Therefore, the active area of the polar cap is assumed to be approximately semicircular.  The calculated Stokes parameters are shown in Fig. 3. They broadly reproduce the general features seen by Manchester et al although the asymmetry in $L$ and $V$ is not present, but could easily be introduced by non-uniformity of the continuum distribution. The spectral index used ($\alpha = 3$) has been estimated from the ATNF fluxes at $400$ and $1400$ MHz.

However, the FWHM  is too large by a factor of $2$.  A possible explanation for this is that the basic assumption of isotropic emission in the turbulence rest frame may not be valid.  In this frame, both ions and protons have considerable velocities parallel and anti-parallel with local flux lines. A unique direction remains and a modest deviation from isotropy would suffice to narrow $f(\chi)$.  Variation of $\gamma$ is limited because the value of $BP^{-2}$ for this pulsar limits the proton Lorentz factor to $\gamma_{p} < 40$ approximately.

\begin{figure}
\includegraphics[trim=20mm 40mm 20mm 60mm, clip, width=84mm]{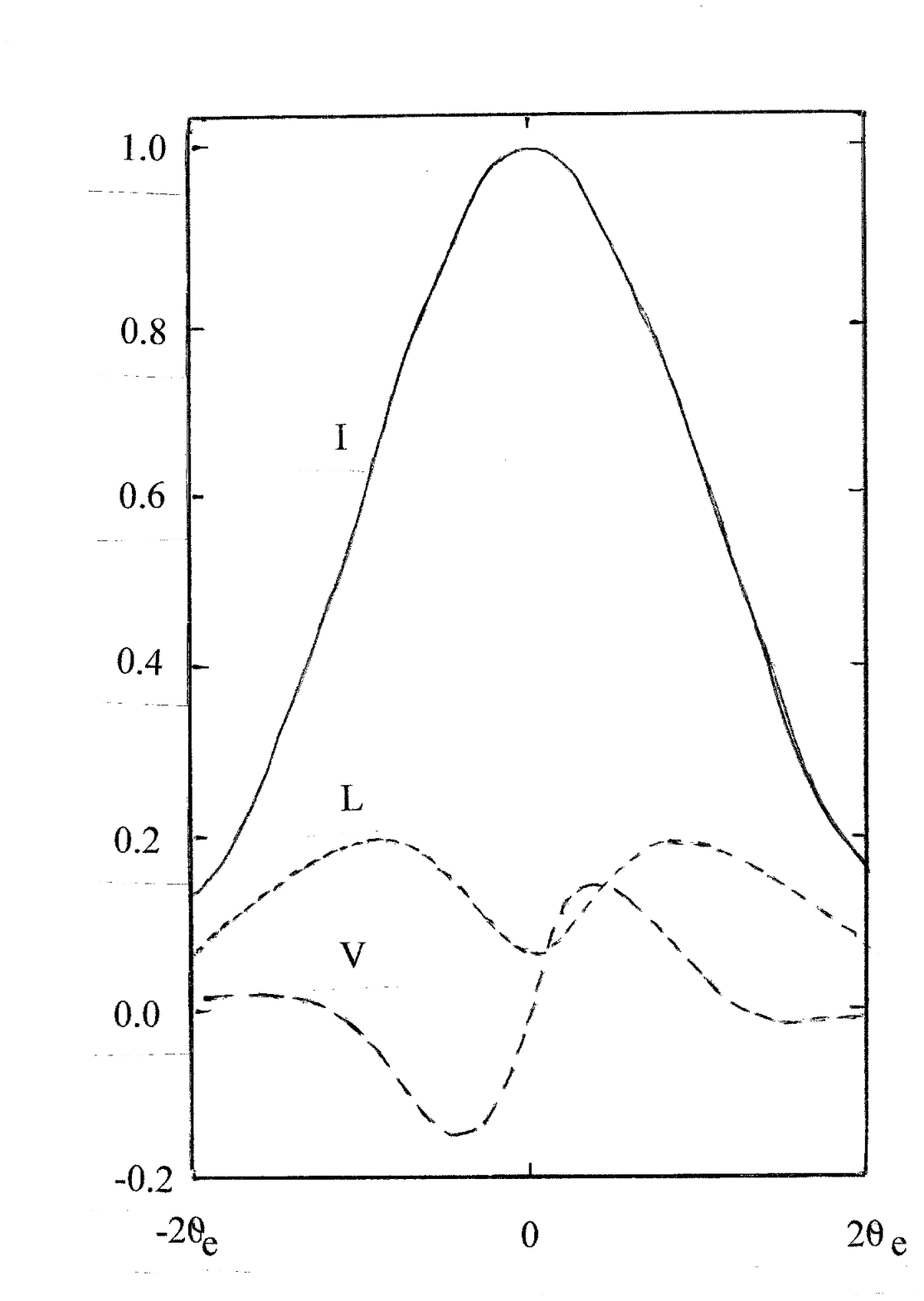}

\caption{Stokes parameters I, $L$, $V$ have been calculated at $0.66$ GHz using the ion-proton model for J2144-3933, and are shown for $\gamma = 30$ and $\eta_{e} = 3$ as functions of longitude as in Fig.1.  The profiles broadly match those observed by Manchester, Han, \& Qiao except for a lack of asymmetry, which could easily be rectified by a small change in the polar-cap intensity distribution at $\eta_{e}$, and for too large a value of the FWHM which exceeds the observed by a factor of $2$. This latter problem is discussed in the text. The observer's line of sight is displaced $d =0.0008$ rad from the magnetic axis and on the side opposite the active area.}

\end{figure}

The polarizations of a set of $9$ pulsars with interpulse emission have been studied by Johnston \& Kramer (2019).  They confirm Desvignes et al in finding that the sign of $V$ depends in a given pulsar on a geometrical factor, the sign of the gradient of the position-angle sweep.  This is as expected in the ion-proton magnetosphere and can be inferred from the form of equation (A12) because both poles can be expected to be described by similar parameters and to have approximately the same O-mode phase lags.

A further interesting feature described by Johnston \& Kramer is the change in sign of $V$ for J0908-4913, between $1.4$ and $3.0$ GHz, at approximately $2.0$ GHz (Kramer \& Johnston 2008).  This is seen in both main pulse and interpulse at the same frequency and is regarded as a problem within the framework of the limiting polarization theory of Beskin \& Philippov.  But equations (5) and (A12) show that the O-mode phase lag $\phi$ is inversely proportional to $\omega$.  Zeros of $V$ occur for $\phi = n\pi$, where $n$ is here an integer.

Let us assume that $\phi = \pi$ at $2.0$ GHz.  Then the product $\gamma\eta_{e}^{2} = 365$ with $\gamma = 20$ and $\eta_{e} = 4.3$ would satisfy, both not unreasonable values.  But if this is correct there should be observable changes in the sign of V at frequencies below $1.4$ GHz for higher values of $n$, that is, at $1.0, 0.7,.......$ GHz.  This assumes that $\eta_{e}$ is frequency-independent, also that $\gamma$ can be assumed constant over the active area of the polar cap, both conditions that are unlikely to be perfectly satisfied.

But $\phi = 2\pi$ at $2.0$ GHz is also a possibility.  In general, equation (5) shows that $V/I \rightarrow 0$ in the limit $\omega \rightarrow \infty$ but with $\omega$ decreasing from that limit $V/I$ alternates in sign through a sequence of identical maximum moduli.  In principle, measurement of $V/I$ as a function of $\omega$ should make possible estimates of $\gamma\eta_{e}^{2}$.

\section{Conclusions}

It is established that limiting polarization in an electron-positron plasma does lead to observable circular polarization in normal pulsars, those for which the light cylinder radius is of the order of $10^{9-10}$ cm (Petrova \& Lyubarskii 2000, Beskin \& Philippov 2012).  But little attention appears to have been paid to the MSP.  The very basic estimates of the limiting polarization radius given in Table 1 are consistent with the work cited for normal pulsars, but also show that limiting polarization in electron-positron plasmas is most unlikely to give an understanding of circular polarization in MSP.  The reason is that the region of limiting polarization is far outside the very small light cylinder radius.  The order of magnitude of $V$ in MSP obtained from Table 1 is broadly consistent with the single result given by Beskin \& Philippov.

However, the ion-proton magnetosphere is also capable of producing the forms of circular polarization that are observed in normal pulsars.  It has the merit of giving an understanding of observed detail, and
values of parameters concerned do not have to be extended beyond reasonable estimates to be consistent with pulsar data.  An aspect of the model which has received almost no detailed attention is the formation and decay of the ion-proton turbulent state.  The longitudinal Langmuir mode has a higher growth rate than the quasi-longitudinal, so that coupling with the radiation field would not be expected to damp the growth of the mode above the polar cap.  Also a two-beam system with different Lorentz factors is not in its lowest energy state at a given total momentum, which would be one of identical Lorentz factors.  Thus radiative decay is natural once the mode becomes turbulent.  We have referred to our lack of understanding of this process in connection with the small FWHM value seen in J2144-3933.  Our assumption that emission in the turbulence rest frame is not isotropic appears to be necessary.

But the ion-proton calculations of Section 3 cannot be applied to the MSP because the O-mode phase lag is proportional to $BP^{-1}$ and would be too small. (The average value for the $24$ MSP observed by Dai et al is $B_{12}P^{-1} \approx 0.05$.)  However, Table 1 indicates that the limiting polarization region lies within the light cylinder for ion-proton MSP and according to the estimates of Lyubarskii \& Petrova should be consistent with values found by Dai et al.  But examination of the profiles published by Dai et al with those of Karastergiou \& Johnston (2006) reveal certain differences.

The MSP I-profiles are, of course, broader and often have more detail in the form of narrow maxima (or components) that appear or disappear as functions of frequency. These indicate discrete active areas of the polar cap near the observer's line of traverse (see Fig. A1).  It is also often the case that in normal pulsars, the $V$-profiles are approximately of the same width as the $I$-profiles, whereas in MSP the $V$-profiles, and to a lesser extent the $L$-profiles, are confined to the same small longitude intervals as the narrow maxima referred to above.  Within these, the $V$-profiles are symmetrical about the maxima with no zeros. An example in the paper of Dai et al is J0613-0200.  But this is precisely the form that limiting polarization analysis predicts for a discrete source (Lyubarskii \& Petrova 1999), whereas the analysis of Section 3 predicts a zero as in J2144-3933 (Fig.3).  This is consistent with our assertion that the changes in field geometry produced by rotation at radii $\eta \sim 5-10$ produce limiting polarization in MSP although in normal pulsars they are negligible.  In both instances an ion-proton magnetosphere is required.

There are many observational similarities which suggest that normal pulsars and MSP have identical magnetospheric composition.  But the problem is, how does an electron-positron plasma form in the MSP or even in normal pulsars such as J2144-3933?  Inverse Compton scattering must occur and has to be considered if there are primary beam electrons (see Hibschman \& Arons 2001).  But however high the ICS photon momentum $K$, its component perpendicular to the local magnetic field must satisfy the condition $K_{\perp}B_{12} \approx 6mc$ to produce a pair. Values of $K$ satisfying this for values of $B$ at most of the order of $10^{8-9}$ G would produce pairs with extremely high Lorentz factors and there would be no possibility that curvature radiation and secondary pair creation could give a plasma of Lorentz factor  no more than $10^{2}$, the value needed for participation in collective longitudinal modes.
Our view is that circular polarization calls into question the canonical view of electron-positron open magnetosphere physics.  Studies of any magnetosphere problem based on it are, for the majority of pulsars, addressed to a non-existent system.

Data availability statement:  the data and algorithms underlying this work will be shared on reasonable request to the corresponding author.

\section*{Acknowledgments}

It is a pleasure to thank Lucy Oswald and Aris Karastergiou for very helpful discussions on pulsar circular polarization and Stig Topp-Jorgensen for the IT arrangements that made working from home possible this summer, also the anonymous referee for interesting comments on phenomenological approaches to pulsar circular polarization.

\appendix
\section{Polar cap details}

The polar-cap charge density is assumed to be positive and the ${\bf B}$-field is described in cylindrical polar coordinates, the magnetic axis being the z-axis.  The polar-cap radius at the neutron-star surface is,
\begin{eqnarray}
u_{0}(1) = \left( \frac{2\pi R^{3}}{1.368cP}\right)^{1/2}
\end{eqnarray}
for neutron-star mass $1.4M_{\odot}$ and radius $R = 1.2 \times10^{6}$ cm (Harding \& Muslimov 2001). Thus the angular radius of the open magnetosphere at radius $\eta_{e}$ is,
\begin{eqnarray}
\theta_{e} = \frac{3u_{0}(1)}{2R}\eta^{1/2} = 0.020P^{-1/2}\eta^{1/2} \hspace{2mm} {\rm rad}
\end{eqnarray}
The plane in Fig. A1 is perpendicular to the magnetic axis and at the source radius $\eta_{e}$, and shows angular deviations from the magnetic axis. It is fixed in the rest-frame of the rotating neutron star. Thus bundles of flux lines bearing turbulent plasma can be represented either as a point at $\theta < \theta_{e}$ and angle $\zeta_{B}$ or by a continuous distribution within the radius $\theta_{e}$.

\begin{figure}
\includegraphics[trim=20mm 45mm 10mm 50mm, clip, width=84mm]{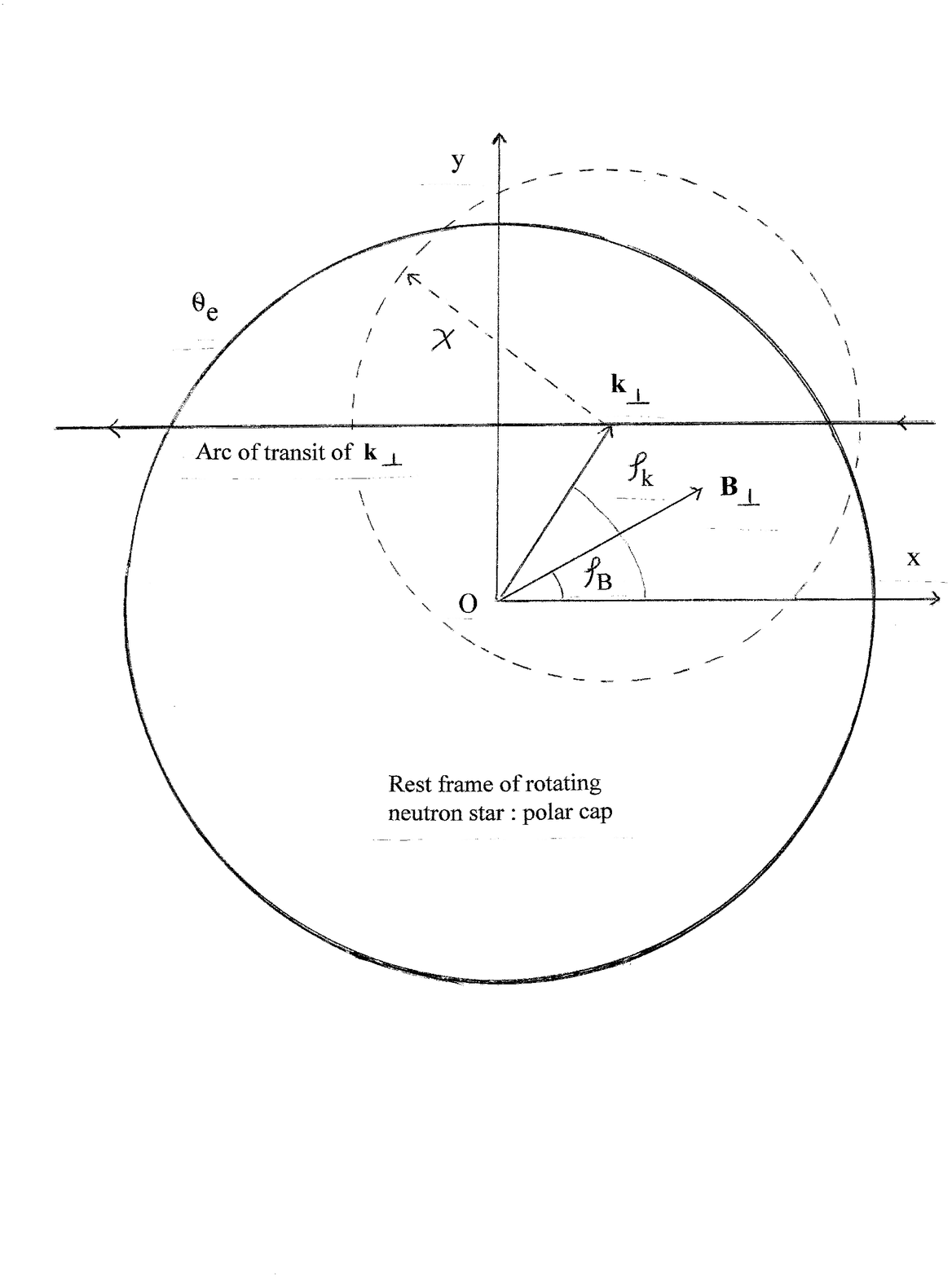}

\caption{This shows angular displacements $0 < \theta < \theta_{e}$ from the magnetic axis O in the emission plane normal to the magnetic axis and at radius $\eta_{e}$ (in units of $R$).  It is in the rest frame of the rotating neutron star and assumes a positive Goldreich-Julian charge density in a dipole field.  Positions in the plane are in cylindrical polar coordinates $(B_{\perp},\zeta_{B})$ and $(k_{\perp}, \zeta_{k})$ where ${\bf k}$ is a unit vector in the line of sight of the observer.  Thus ${\bf k}_{\perp}$ moves across the diagram from the right, on the arc of transit, here rectilinear.  Owing to radiation being emitted at a finite angle $\chi$ to the local field and particle direction, Stokes parameters at a specific ${\bf k}_{\perp}$ are found by summation over those for elements at all points on the emission surface.  The displacement of the line of traverse from the magnetic axis is $d = k_{\perp}\sin \zeta_{k}$.  Small-angle approximations are made for vector products.}

\end{figure}

The line of sight is the unit vector ${\bf k}$ with cartesian coordinates
$(k_{\perp}\cos \zeta_{k}, k_{\perp}\sin \zeta_{k})$. It moves from right to left on the observer's arc of traverse (here rectilinear).  Small angle approximations are used throughout for $\theta$ and $\chi$.  The circle centred on ${\bf k}_{\perp}$ represents notionally the fact that radiation from flux lines at an angle $\chi$ to ${\bf k}_{\perp}$ can reach the observer as determined by the function $f(\chi)$ defined by equation (3).

The plasma O and E-mode position angles can be found at any point in the plane for a given source linear polarization.  Our assumption, prompted by the RVM, is that the radiation is polarized with position angle given by the coordinates ${\bf s} = (-\cos \zeta_{B}, -\sin \zeta_{B})$, that is, locally parallel with ${\bf B}_{\perp}$.  The E-mode electric field components at $\eta_{e}$, unit vector ${\bf q}_{E}$ are,
\begin{eqnarray}
q_{Ex} = A(-k_{\perp}\sin \zeta_{k} + k\theta\sin \zeta_{B}),\nonumber\\
q_{Ey} = A(-k\theta\cos \zeta_{B} + k_{\perp}\cos\zeta_{k})
\end{eqnarray}
where,
\begin{eqnarray}
A =(k_{\perp}^{2} + k^{2}\theta^{2} - 2k\theta k_{\perp}
\cos(\zeta_{k} - \zeta_{B}))^{-1/2},
\end{eqnarray}
and for the O-mode,
\begin{eqnarray}
q_{Ox} = A(k\theta\cos\zeta_{B} - k_{\perp}\cos\zeta_{k})  \nonumber \\
q_{Oy} = A(k\theta\sin\zeta_{B} - k_{\perp}\sin\zeta_{k})
\end{eqnarray}
The electric field components at $\eta > \eta_{e}$ are then,
\begin{eqnarray}
E_{x} = (E_{xE} + E_{xO}\exp(-i\phi))\exp(i\omega t)
\end{eqnarray}
and,
\begin{eqnarray}
E_{y} = (E_{yE} + E_{yO}\exp(-i\phi))\exp(i\omega t),
\end{eqnarray}
with $E_{xE} = {\bf s}\cdot{\bf q}_{E} q_{Ex}$ and $E_{xO} = {\bf s}\cdot{\bf q}_{O}q_{Ox}$ etc.

The Stokes parameters are,
\begin{eqnarray}
I = E^{2}_{xE} + E^{2}_{xO} +E^{2}_{yE} + E^{2}_{yO}  \nonumber \\
 + 2(E_{xE}E_{xO} + E_{yE}E_{yO})\cos\phi,
\end{eqnarray}
\begin{eqnarray}
Q = E^{2}_{xE} + E^{2}_{xO} - E^{2}_{yE} - E^{2}_{yO}   \nonumber   \\
 + 2(E_{xE}E_{xO} - E_{yE}E_{yO})\cos\phi,
\end{eqnarray}
\begin{eqnarray}
U = 2(E_{xE}E_{yE} + E_{xO}E_{yO})  \nonumber   \\
 + 2(E_{xE}E_{yO} + E_{yE}E_{xO})\cos\phi,
\end{eqnarray}
\begin{eqnarray}
V = -2(E_{xE}E_{yO} - E_{yE}E_{xO})\sin\phi.
\end{eqnarray}
The explicit expression for Stokes $V$ is then,
\begin{eqnarray}
\frac{{V}}{{I}} = 2\frac{(-k\theta + k_{\perp}\cos(\zeta_{k}
 - \zeta_{B}))k_{\perp}\sin(\zeta_{k} - \zeta_{B})\sin\phi}
 {k^{2}\theta^{2} + k^{2}_{\perp} -2k\theta k_{\perp}
 \cos(\zeta_{k} - \zeta_{B})}.
 \end{eqnarray}

\bsp

\label{lastpage}

\end{document}